\documentclass[pra,aps,amsmath,amssymb,notitlepage,twocolumn,floatfix,superscriptaddress]{revtex4-1}
\usepackage{amsthm}
\usepackage{amsfonts}
\usepackage{siunitx}
\usepackage{amsmath}
\usepackage{amssymb}
\usepackage{graphicx}
\usepackage{verbatim}
\usepackage[colorlinks]{hyperref}
\usepackage{tikz}
\usepackage{pgfplots}
\usepackage{braket}

\usepackage{enumitem} 

\definecolor{linkcolor}{RGB}{0,83,166}
\hypersetup{
    colorlinks = true,
    allcolors = {linkcolor}
}

\pgfplotsset{compat=newest}
\usetikzlibrary{plotmarks}
\usepgfplotslibrary{groupplots}

\pgfdeclarelayer{foreground}
\pgfsetlayers{main,foreground}

\newlength\figureheight
\newlength\figurewidth

\renewcommand{\epsilon}{\varepsilon}

\pgfplotsset{every axis/.append style={
    axis line style=thick,
  }
}

\begin{document}
\title{Comment on ``Scaling advantages of all-to-all connectivity in physical annealers:  the Coherent Ising Machine vs D-Wave 2000Q'' }
\author{Catherine C. McGeoch}\email[]{cmcgeoch@dwavesys.com}
\author{William Bernoudy}
\author{James King}
\affiliation{D-Wave Systems Inc, Burnaby B.C.} 
\date{\today}

\begin{abstract}
 A recent paper compares performance of a D-Wave 2000Q system to two versions of the 
coherent Ising machine (CIM).  This comment points out some flaws in the analysis,
which undermine several of the conclusions drawn in the paper.  
\end{abstract}

\maketitle


\section{Introduction} 
A recent paper by Hamerly et al. [1] (see also [2]) considers the impact of connectivity on physical 
annealing systems.  Starting from a premise that connectivity has a significant impact on 
performance,  the authors describe results from 
experiments performed on a D-Wave 2000Q annealing-based quantum processor to two versions 
of the coherent Ising machine (CIM).   The former  has relatively sparse physical connectivity 
among its qubits;  the latter has all-to-all connectivity in its FPGA subsystem but no physical 
connectivity among the elements in its optical subsystem.   

Test inputs from two problem classes, here denoted SK (Sherrington-Kirkpatrick) and 
MC (Max Cut), are used.  The performance metric of interest is time-to-solution (TTS), calculated from 
time to find a single solution and the probability that a given solution is optimal.  
The take-away messages from this paper are summarized in this excerpt from the abstract:  

\begin{quote}
{\bf Abstract:}  We demonstrate an exponential $O(e^{-O(N^2))})$ penalty in performance 
for the D-Wave quantum annealer relative to coherent Ising machines when solving Ising problems on 
dense graphs, which is attributable to the differences in internal connectivity between the machines.  This leads 
to a several-orders-of-magnitude 
time-to-solution difference between coherent Ising machines and the D-Wave system for problems with 
over 50 vertices.  Our results provide 
strong experimental support to efforts  to increase the connectivity of physical annealers.  
\end{quote} 
The starting premise, that physical connectivity plays a role in performance of these systems, has been well discussed in the   
literature and is not in dispute.  The conclusion, in support of increased connectivity in physical  annealers,
aligns with ongoing  work at D-Wave and elsewhere to build next-generation quantum systems with denser connection 
topologies.  However, the remaining claims in this excerpt are based on flawed analysis and 
faulty reasoning about what conclusions can legitimately be drawn from the data:

\begin{enumerate} 
\item The ``exponential $O(e^{-O(N^2))})$'' performance gap is based on two regression models that 
are meant to describe scaling of the algorithms realized by the 2000Q processor and the CIM.   However,  the 
authors have fit their models incorrectly, to data that does not correspond to the true scaling of these 
algorithms. The resulting curves {\em overestimate} scaling of the 2000Q processor and {\em underestimate} 
scaling of the CIM. 

\item The ``several-orders-of-magnitude'' difference in TTS is based on extrapolating those incorrect curves
to large problems well outside the range of tests.  Looking at the {\em measured} data,  the CIM appears to be 
about 100x and 8000x faster on SK and dense MC inputs, respectively.   However, an apples-to-apples 
analysis of their data (treating post processing the same way for both systems) shows much smaller gaps:  the 
CIM is about 5x faster on SK and 364x faster on dense MC inputs.  That gap might have been further reduced, 
or perhaps eliminated, if the tests had incorporated performance-tuning features available on 
the 2000Q processor. 

\end{enumerate} 
The following sections discuss these issues in more detail.  

First, however,  we remark that even if the CIM is faster on dense SK and MC inputs (and the 2000Q 
system is faster on sparse inputs, as described in [1,3]),  there is nothing in the paper to suggest that 
this result is representative of performance on inputs that matter.   

That is, many classical optimization algorithms,  such as simulated annealing and the CIM,  work by 
stepping through a large space of possible solutions one at a time.  Some inputs are {\em combinatorially easy,}   
requiring just a few steps and time scales in fractions of seconds to be solved;  
some inputs are {\em combinatorially hard,}  requiring an enormous number of steps and time 
scales in hours, weeks, or years, to be solved.   

D-Wave quantum processors work under a completely different paradigm:  instead of 
of stepping through the solution space, they exploit quantum properties to find 
shortcut routes to optimal solutions.  The inputs that matter to quantum performance evaluation are the 
combinatorially hard ones:   if a given set of inputs is easy enough to be solved quickly by classical  methods,  
there is no room for improvement and nothing to be gained by using the quantum approach.  
  
While it is easy to see how graph density would affect computation time per step,  the more important 
question  is whether it affects the total number of steps required in the classical computation.   This 
question is not addressed in the Hamerley et al. paper,  but the data does give hints that the problems are 
combinatorially easy.  For example, the CIM apparently never needs more than $1000$ steps to search 
a solution space of size $2^{150}$;  and on SK problems, while input size increases by one order of magnitude, 
computation time for the CIM grows by less than an order of magnitude.  Furthermore, results by King et al. [4] 
show that these problems can be solved in less than 10 milliseconds using a GPU implementation 
of the CIM algorithm,  which runs roughly 20 times faster than the CIM.    

More generally,  it is known that higher density doesn't necessarily make NP-hard graph 
problems {combinatorially} harder.  In many problem domains it has been 
demonstrated that dense problems can be solved {more efficiently} than sparse problems (see e.g. [5]). 
Furthermore,  it is known Max Cut problems on dense graphs can be approximately solved by efficient greedy 
methods [6];  without going into technical details,  this would translate to efficient exact solution methods 
on finite-sized inputs.  

In summary:  the scaling analysis does not correspond to the true scaling of the two algorithms under 
study;  the runtime analysis does not correspond to performance that would have been observed under 
balanced test conditions;  and there is no evidence that the inputs selected for testing 
are hard enough to support conclusions about performance in general.   For these reasons, 
the conclusions drawn in the Hamerly et al. paper do not give an accurate picture of the 
true performance landscape for these systems. 

\section{Faulty Regression Analysis} 
Annealing-based optimization algorithms have an {anneal time} parameter $t$ that controls the 
amount of computational effort spent solving a given input.  For classical algorithms, $t$ determines the 
number of steps needed to explore the solution space, inspecting one solution at a time.  For an annealing-based 
quantum processor, $t$ corresponds to elapsed time in an analog computation, which must be above a 
certain threshold to make successful outcomes possible. 

In both cases, increasing anneal time is associated with increasing success probability.  
Let $t_{p}(N)$ denote the minimum anneal time sufficient to achieve some target success 
probability $p$, as a function of problem size $N$.  The amount of computational work 
performed by an annealing-based algorithm corresponds to how $t_{p}(N)$ scales with $N$.  

Fig. 2 of Ref. [1] shows four regression curves corresponding to four values of $t$, which are fit to $p$ 
using the regression model $P = exp(-(N/N_0^{DW})^2)$, where $N_0^{DW}$ is a slowly-increasing function of $t$.    

This model is used to describe scaling of the D-Wave 2000Q processor,  but note that the 
curve is fit to the wrong data:  the scaling of $p$ for fixed $t$ is not the same as the  
scaling of $t$ for fixed $p$.  Given a constant amount of effort to solve the 
problem,  one would expect success probabilities to plummet to 0 as $N$ grows;   but this is 
completely independent of how much effort is needed to maintain a given success probability.  

This difference is explained by R{\o}nnow et al. [7], who show that  the ``true scaling''  
curve for an annealing-based 
algorithm corresponds to the {\em lower envelope} of the sequence of fixed-anneal time 
curves.  Fig. 1 of  ref.  [7]  
shows true scaling curves for two classical solvers, which approach different asymptotes, 
from different directions (convex lower bound vs. concave upper bound) than do the fixed-anneal curves. 

Furthermore,  R{\o}nnow et al.  show how a solver that does too much work at small $N$ 
can exhibit ``fake speedup''  with data that appears flatter than the true scaling curve.  
Fig. 2(c) of [1] shows that the CIM curve is nearly horizontal and 
does not intersect the origin, a strong indication that fake speedup is present.   If so, the 
regression model used in the paper {underestimates} true scaling for the CIM.    

The combination of regression models that overestimate 2000Q scaling and underestimate CIM scaling 
produces the performance gap claimed in the abstract, which greatly overestimates the true gap.  
From the data presented in the paper it is impossible to determine conclusively whether a scaling 
gap exists at all, since the lower envelope curve in Fig. 2 is too small to support trustworthy data analysis.   

\section{Imbalance in Runtime Comparisons}  
Putting aside the flawed extrapolation and considering runtime performance that was actually measured,  Fig. 2(c) suggests 
that the CIM is about 100x faster than the D-Wave 2000Q system on SK instances, and Fig S6 shows the CIM to be 
about 8000x faster on dense MC instances at $N=60, N=45$ respectively.  

However, Appendix S2 mentions that the CIM TTS calculations include  a post-processing step that selects the ``best batch''  of results 
from several trials:   ``This is the success probability we could expect from a well-engineered CIM 
where the optical phase, 
pump power, and other optical degrees of freedom  have been sufficiently stabilized.''   Fig S5 
shows that success probabilities for raw versus postprocessed data are near $p_{raw}= .1,  p_{post}=.9$ at the problem sizes of interest:  this translates to about a 22x speedup in  TTS from postprocessing.  Comparing {\em raw} CIM performance to 
raw 2000Q performance,  the CIM is about 5x faster on SK inputs and about 364x faster on dense MC inputs.  

It would be interesting to know whether 2000Q performance might have improved if this ``best batch''  approach had been 
applied, say, to batches of results from each spin reversal transform, or to batches from different problem embeddings.   It would also be interesting to know if using any of the built-in tools for improving raw results on the 2000Q system (such as 
modifying the anneal path,  using the extended J-range or virtual graph capabilities, or postprocessing) might have yielded faster TTS. 
 
\section{Conclusion}

The hazards of drawing conclusions about computational performance, based on 
extrapolation from unvalidated regression models, are widely 
understood (e.g. see [8]).   Extrapolating performance results beyond the range of inputs tested is 
equally hazardous.  These hazards are exacerbated in this time of rapid 
development of quantum computer technologies,  when 
performance models are weak or nonexistent,  and the data itself changes with 
each new processor generation. 

\section*{Bibliography} 
{\footnotesize

\begin{enumerate}[label={[\arabic*]}]
\item  Hamerly et al. ``Scaling advantages of all-to-all connectivity in physical annealers: the Coherent Ising Machine vs. D-Wave 2000Q,'' 
arXiv 1805.05217v1, 14 May 2018.   

\item  Hamerly et al., ``Quantum vs. Optical Annealing: Benchmarking the OPO Ising Machine and D-Wave,''  
in {\em Conference on Lasers and Electro-Optics}, OSA Technical Digest, Optical Society of America, 2018. 

\item R. M. Hamerly,  ``Quantum vs. Optical Annealing:  the Coherent Ising Machine and DWAVE,''  presentation at AQC conference, 2017.  


\item King et al., ``Emulating the coherent Ising machine with a mean-field algorithm,''  arXiv:1806.08422v1,  21 June 2018.  

\item Banderier et al., ``Average case analysis of NP-complete problems: Maximum independent set and exhaustive search 
algorithms,''  {\em Proceedings of AofA'09}, 2009. 

\item Mathieu and Schudy, ``Yet another algorithm for dense max cut: go greedy,''  Proceedings of SODA'O8, pp. 176-182, 2008.   

\item  R\o nnow et al., ``Defining and detecting quantum  speedup,'' {\em Science} 345, 420, 2014.   

\item David Bailey, ``Twelve ways to fool the masses when giving performance results on parallel computers,''  {\em Supercomputing Review},
pp. 54-55,  Aug 1991.   See also the updated version,  http://www.davidhbailey.com/dhbtalks/dhb-12ways-2013.pdf.   


\end{enumerate} 
} 

\bibliography{cim}
\appendix

\end{document}